\documentclass{PoS}

\usepackage{amsmath}
\usepackage{amsfonts}
\usepackage{mathtools}
\usepackage{graphicx}
\usepackage[utf8]{inputenc}
\newcommand{\fsl}[1]{\ensuremath{\mathrlap{\!\not{\phantom{#1}}}#1}}% \fsl{<symbol>}
\newcommand{\figWidth}{0.464}
\title{Masses, decay constants and electromagnetic form-factors with twisted boundary conditions} 

\ShortTitle{Masses, decay constants and em form-factors with twisted boundary conditions}

\author{\speaker{Johan Relefors}%
  \\
  Theoretical High Energy Physics, Lund University\\
  E-mail: \email{johan.relefors@thep.lu.se}}

\author{Johan Bijnens\\
  Theoretical High Energy Physics, Lund University\\
  E-mail: \email{bijnens@thep.lu.se}}

\abstract{We discuss some of the effects of twisted boundary conditions in finite volume using continuum SU(3) Chiral Perturbation Theory. We point out how broken cubic symmetry affects the definitions of quantities such as form-factors. Using the $\pi^+$ as an example, we give one loop results for the mass, decay constants and electromagnetic form-factor and illustrate how the relevant Ward identities are satisfied.}

\FullConference{The 8th  International Workshop on Chiral Dynamics \\
  29 June 2015 - 03 July 2015\\
  Pisa, Italy}

\FullConference{{\rm To be published in the proceedings of:\\}
The 8th  International Workshop on Chiral Dynamics \\
  29 June 2015 - 03 July 2015\\
  Pisa, Italy}
\begin{document}

\section{Introduction}

Lattice QCD simulations necessarily take place in a finite volume. This gives unphysical effects that must be compensated for. For example, using periodic boundary conditions removes the actual boundary in order to eliminate boundary effects but introduces conditions on the momenta available in the simulation. At a lattice size of 5 fm this means that the lowest lying non-zero momentum is
\begin{align}
  p(1) \approx 250 ~\mathrm{MeV}.
\end{align}
Depending on the process under study this can be considered large, and furthermore there are situations where being able to continuously vary momenta is advantageous. Examples include simulations for the charge radius of the pion \cite{Brandt:2013dua}, hadronic vacuum polarization for $\mu$ g-2 \cite{Aubin:2013daa} and $K_{l3}$ decays close to zero momentum which is relevant for determination of $V_{us}$ \cite{Boyle:2015hfa}. Twisted boundary conditions, which are introduced in section \ref{seq:twisted}, is a way of continuously varying momenta in lattice QCD \cite{Bedaque,twisted,twisted2}.

In this talk we investigate some effects of twisted boundary conditions using continuum SU(3) Chiral Perturbation Theory in finite volume at one loop \cite{Weinberg,GL1,GL2,GL4}. In particular we show how Ward identities are satisfied in a nontrivial way and how this relates to meson masses. Full results for all calculations will not be presented since they are already available in \cite{Bijnens:2014yya}. We focus instead on the principal differences between the case of periodic boundary conditions and twisted boundary conditions using results for $\pi^+$ to illustrate. See also \cite{JT} for discussions on some of the issues covered here.

\section{Twisted boundary conditions}
\label{seq:twisted}
Twisted boundary conditions are defined through the relation
\begin{align}
  \phi(x^i+L) = \phi(x^i)\exp(i\theta^i).
\end{align}
where we have assumed that the field is confined to a cube where all sides are of equal length L and with $i$ running over spatial dimensions. Expanding both sides in Fourier modes gives
\begin{align}
  \int dp^i \phi(p^i)\exp({ip^i(x^i+L)}) = \int dp^i \phi(p^i)\exp({i(p^ix^i+\theta)}) \Rightarrow p^i = p^i(n^i) = \frac{2\pi}{L}n^i+\frac{\theta^i}{L},\, n\in\mathbb(Z).
\end{align}
By varying the twist angle, $\theta$, it is possible to vary the momentum of the field continuously. There are two more important observations we want to point out. First, twisted boundary conditions break the cubic symmetry of the lattice which means that e.g. form-factors are not as constrained. Second, available momentum states are no longer symmetric around zero. This has the consequence that tadpole integrals are no longer zero and, more generally, that there are new terms in loop integrals which reflects the loss of cubic symmetry. These two related observations were the starting point of our calculations.

\subsection{Twist for quarks}
The Dirac term in the QCD lagrangian is
\begin{align}
  \mathcal{L} = \bar {q}(i\fsl{D}- m)q
\end{align}
where $\bar {q}$ and $q$ are Dirac spinors and also vectors in flavour space, $D$ is the covariant derivative in QCD and $m$ is the flavour diagonal mass matrix. We now define twisted boundary conditions for each quark flavour separately as
\begin{align}
  q_f(x^i+L) = \exp(i\theta^i_f)q_f(x^i).
\end{align}
This definition makes it clear that each flavour and direction can have its own twist angle $\theta^i_f$. To make contact with methods developed for periodic boundary conditions it is practical to undo the twist of the quarks using a field redefinition \cite{twisted}. This redefinition can be seen as a local $U(n)_V$ flavour transformation for $n$ flavours, using only the Cartan subalgebra. It is however not to be confused with a gauge transformation in color space. Collecting the twist angles for the flavour $f$ in a three vector $\vec \theta _f$ we define
\begin{align}
  \tilde {q}_f(x) = q_f(x)\exp\left(-i\vec \theta_f \cdot\frac{\vec x}{L}\right)\,.
\end{align}
The field $\tilde {q}$ satisfies periodic boundary conditions
\begin{align}
  \tilde {q}(x^i+L) = \tilde {q}(x^i).
\end{align}
The lagrangian for the field $\tilde {q}$ is then
\begin{align}
  \mathcal{L} = \bar {\tilde {q}}(i\fsl{D} + i\frac{\fsl\Theta}{L}- m)q
\end{align}
where we've collected the twist angles in a flavour diagonal matrix $\Theta^\mu$ where the entries are four vectors $\theta^\mu_f = (0,\vec\theta_f)$. We can now interpret the term $\bar{\tilde {q}} i\fsl\Theta/L\tilde {q}$ as coming from a constant background field. Note that it interacts differently with the different flavours since it is proportional to the twist angles. Note that the quarks, $\tilde {q}$, have momenta $\vec k = \frac{2\pi}{L}\vec n$.

\subsection{Twist for mesons}

A meson with the flavour structure $\bar {q}^\prime q$ has the twist angle
\begin{align}
  \phi_{\bar {q}^\prime q}(x^i+L) = \exp(i(\theta^i_q-\theta^i_{q^\prime}))\phi_{\bar {q}^\prime q}(x^i) \equiv \exp(i\theta^i_{\bar {q}^\prime q})\phi_{\bar {q}^\prime q}(x^i).
\end{align}
We will use the three- and four-vector notation for meson twists as we did above for the quarks and we will label the mesons $\phi_{\bar {q}^\prime q}$ with $\pi^\pm,...$. Note that even though the flavour diagonal mesons have twist angle zero, they are affected through mixing in the cases where twisted boundary conditions break isospin symmetry. The broken isospin symmetry affects the decay constants at NLO while the effect for the masses is NNLO \cite{ABT2}. For charged particles charge conjugation is broken. This happens since particle and anti-particle have opposite twist angles. A particle with spatial momentum $\vec p$ corresponds to an anti-particle with spatial momentum $-\vec p$.

\subsection{Calculating Feynman diagrams with twisted boundary conditions}
\label{Calculating}
We will now introduce the integral notation used in the rest of the talk, for a more complete treatment see \cite{Becirevic,sunsetfiniteV,Bijnens:2014yya}. We label tadpole integrals with A according to
\begin{align}
  A^{\{\, ,\mu,\mu\nu\}}(m^2,n) = \frac{1}{i}\int\frac{d^dk}{(2\pi)^d}\frac{\{1,k^\mu,k^{\mu\nu}\}}{(k^2-m^2)^n}
\end{align}
and two propagator integrals with B according to
\begin{align}
  B^{\{\, ,\mu,\mu\nu\}}(m^2_1,m_2^2,n_1,n_2) = \frac{1}{i}\int\frac{d^dk}{(2\pi)^d}\frac{\{1,k^\mu,k^{\mu\nu}\}}{(k^2-m^2_1)^{n_1}((q-p)^2-m_2^2)^{n_2}}.
\end{align}

When spatial momenta are quantized, the spatial part of loop integrals are turned into sums over spatial momenta as
\begin{align}
  \int \frac{d^dk_\phi}{(2\pi)^d} \rightarrow  \int_V \frac{d^dk_\phi}{(2\pi)^d} = \int \frac{d^{d-3}k_\phi}{(2\pi)^{d-3}}\frac{1}{L^3}\sum_{\vec n\in\mathbb{Z}^3,\, \vec k=(2\pi \vec n + \vec\theta^\phi)/L}.
\end{align}
In \cite{twisted} it is explained how all propagators in a Feynman diagram end up with allowed momenta.

The sum of course modifies the numerics of each integral but also explicitly breaks cubic invariance. This means that the finite volume parts of loop integrals, labeled by $A^V$ and $B^V$, do not decompose into Lorentz structures in the usual way. Most notably 
\begin{align}
  A^{\mu}(m^2_\phi,n) = \frac{1}{i}\int\frac{d^dk}{(2\pi)^d}\frac{k^\mu}{(k^2-m^2_\phi)^n} \neq 0.
\end{align}
so that the odd tadpole integral is non-zero even though there is no Lorentz structure available. This happens due to the fact that momenta are not symmetric around zero so that the usual cancellation does not occur. Other decompositions that we use in our results are, using $B^{\{\, ,\mu,\mu\nu\}}(m_1^2,m_2^2,1,1) = B^{\{\, ,\mu,\mu\nu\}}(m_1^1,m_2^2)$,
\begin{align}
  B^{V\mu}(m_1^2,m_2^2) &= q^\mu B_1^V(m_1^2,m_2^2,q)+B_2^{V\mu}(m_1^2,m_2^2,q)\nonumber\\
  B^{V\mu\nu}(m_1^2,m_2^2) &= q^\mu q^\nu B_{21}^V(m_1^2,m_2^2,q)+g^{\mu\nu} B_{22}^V(m_1^2,m_2^2,q)+B_{23}^{V\mu\nu}(m_1^2,m_2^2,q)\nonumber\\
\end{align}
For the untwisted case the integral $B_2^{V\mu} = 0$ while $B_{23}^{V\mu\nu}$ is zero in infinite volume. Note that the integrals on the right hand side depend on the components of $q$, not only on $q^2$ as would be the case in infinite volume. Note that all integrals containing twisted mesons depend on twist angles in the finite volume results. It is therefore important to specify whether there is a $\pi^+$ or a $\pi^-$ in a loop and similarly for other particle anti-particle pairs.

When presenting our results we will use the notation
\begin{align}
  \Delta^V X = X(V) - X(\infty)
\end{align}
where X is the object under discussion. We call $\Delta^V X$ the finite volume correction.

\section{Masses}
In order to discuss masses, we first need to define what we mean by mass. Our definition is
\begin{align}
  m^2 = E^2-\vec p^2
\end{align}
for fixed spatial momentum $\vec p$. This will result in a momentum dependent mass $m=m(\vec p)$ due to the broken cubic symmetry. Our definition should be contrasted with the definition \cite{JT}
\begin{align}
  m^2 = E^2 - (\vec p + \vec K)^2,
\end{align}
where $\vec K$ is a NLO correction to $\vec p$ which removes the momentum dependence. The two definitions differ simply by moving around parts of the loop corrections. Note that the shift in momentum is not universal \cite{Bijnens:2014yya}. With this we mean that, if we define the decay constant as
\begin{align}
  \left< 0|A^M_\mu|M(p)\right> = i\sqrt 2 F_M \left(\vec p_\mu+\vec K^\prime_\mu\right),
\end{align}
in order to absorb the momentum dependence of the decay constant we would need to renormalize the momentum differently, that is $\vec K^\prime \neq \vec K$. Note also that it is not possible to find any $\vec K$ such that the electromagnetic form-factor can be written as a single $f_+$ with rescaled momentum.

Now that we have defined mass, we quote the one loop result for the pion mass to illustrate the differences that come up when using twisted boundary conditions;
\begin{align}
  \Delta^V\! m_{\pi^\pm}^2 &= 
  \frac{\pm p^\mu}{F_0^2}
  [- 2 A^V_\mu(m_{\pi^+}^2)-  A^V_\mu(m_{K^+}^2) +  A^V_\mu(m_{K^0}^2)]\nonumber\\
  &+\frac{m_\pi^2}{F_0^2}
  \left( - \frac{1}{2} A^V(m_{\pi^0}^2) + \frac{1}{6}A^V(m_{\eta}^2)\right).
\end{align}
The factor $\pm p^\mu$ in the first row introduces momentum dependence to the mass. All integrals in the first row vanish as the twist angles go to zero. Note that the mass shifts for a $\pi^+$ with momenta $\vec p$ and a $\pi^-$ with momenta $-\vec p$ are the same which is consistent with that charge conjugation should be defined with a minus sign for momenta. The corrections for flavour neutral mesons remain momentum independent.

\section{Electromagentic form-factor and Ward identities}
We define the electromagnetic form-factor by
\begin{align}
  \left<M^\prime(p^\prime)|j_\mu^I|M(p)\right> &= f_{IMM^\prime  \mu}
  \nonumber\\
  &= f_{IMM^\prime  +} (p_\mu+p_\mu^\prime) 
  + f_{IMM^\prime  -} (p_\mu-p_\mu^\prime) 
  +  h_{IMM^\prime  \mu}\,.
\end{align}
where $M$ and $M^\prime$ are mesons and $j^\mu$ is the electromagnetic current. We split up the form-factor into three parts with different momentum dependence. The split is not unique but comes naturally when calculating the finite volume corrections. As long as we only look at the electromagnetic current, the Ward identity ensures that $f_{IMM^\prime  -}$ and $h_{IMM^\prime  \mu}$ go to zero when the twist angles go to zero. For the charged pions the one loop finite volume corrections are
\begin{align}
  \Delta^V\! f_{em \pi^\pm +} &=\frac{\pm1}{F_0^2}\left( 2 H^V(m_{\pi^+}^2,m_{\pi^-}^2,q)
    + H^V(m_{K^+}^2,m_{K^-}^2,q)\right)\nonumber\\
  \Delta^V\! f_{em \pi^+(\pi^-) -} &=
  \frac{p^{\prime\nu}(-p^\nu)}{F_0^2}\left(2 B^V_{2\nu}(m_{\pi^+}^2,m_{\pi^-}^2,q)
    + B^V_{2\nu}(m_{K^+}^2,m_{K^-}^2,q)\right)\nonumber\\
  \Delta^V\! h_{em \pi^\pm\mu} &=
  \frac{1}{F_0^2}\bigg( 2A^V_\mu(m_{\pi^+}^2)+A^V_\mu(m_{K^+}^2)-A^V_\mu(m_{K^0}^2)
  \nonumber\\*
  &
  +q^2B^V_{2\mu}(m_{\pi^+}^2,m_{\pi^-}^2,q)
  +\frac{q^2}{2}B^V_{2\mu}(m_{K^+}^2,m_{K^-}^2,q)
  \nonumber\\*
  &
  \mp(p+p^\prime)^\nu\left(2B^V_{23\mu\nu}(m_{\pi^+}^2,m_{\pi^-}^2,q)
    +B^V_{23\mu\nu}(m_{K^+}^2,m_{K^-}^2,q)\right)\bigg)\,,
\end{align}
where, apart from the integrals defined in section \ref{Calculating}, we have used a generalization of $\mathcal{H}$ found in \cite{GL3,BT}
\begin{align}
  H^V(m_1^2,m_2^2,q) = \frac{1}{4} A^V(m_1^2) + \frac{1}{4}A^V(m_2^2)-B_{22}^V(m_1^2,m_2^2,q).
\end{align}

The Ward identity
\begin{align}
  (p_\mu-p_\mu^\prime) f_{IMM^\prime  \mu} = 0
\end{align}
is still satisfied. This does take some extra care as we show below. Expanding $f_{IMM^\prime  \mu}$ we get
\begin{align}
  0 = q_\mu\left<M^\prime(p^\prime)|j_\mu^I|M(p)\right> = f_{IMM^\prime  +} (p^2-{p^\prime}^2) 
  + f_{IMM^\prime  -} q^2 
  +  h_{IMM^\prime  \mu}q_\mu
\end{align}
where $q_\mu = (p_\mu-p_\mu^\prime)$. In the infinite volume case for $M=M^\prime$ we always have $(p^2-{p^\prime}^2)=0$ on shell to all orders. In the case of twisted boundary conditions this is no longer true. Instead we have, using $\pi^+$ as an example,
\begin{align}
  \left(p^2-{p^\prime}^2\right)_{NLO} = \frac{p^\mu-p^{\prime\mu}}{F_0^2}[- 2 A^V_\mu(m_{\pi^+}^2)-  A^V_\mu(m_{K^+}^2) +  A^V_\mu(m_{K^0}^2)]
\end{align}
which gives a non-zero contribution when multiplied with the leading order part of $f_{IMM^\prime  +}$. This contribution cancels parts of $h_{IMM^\prime  \mu}q_\mu$. The reminaing terms cancel between $f_{IMM^\prime  -} q^2$ and $h_{IMM^\prime  \mu}q_\mu$ using some integral identities \cite{Bijnens:2014yya}.

An issue which is not solved by using twisted boundary conditions is that for $M=M^\prime$ there is no way to continuously vary the momentum in the current since $p^i-p{^i\prime} = \frac{2\pi}{L}(n^i-n{^i\prime})$. In order to address this case it has been proposed in \cite{UKQCD1} that broken isospin symmetry should be used to relate matrix elements through
\begin{align}
  \left<\pi^+(p^\prime)|\bar{u}\gamma_\mu u |\pi^+(p)\right> = -\frac{1}{\sqrt2}\left<\pi^0(p^\prime)|\bar{d}\gamma_\mu u |\pi^+(p)\right>.
\end{align}
Note that in this way contributions from the $\bar s\gamma_\mu s$ part of the current are missed.

\begin{figure}[t]
  \begin{center}
    \includegraphics[width=\figWidth\textwidth]{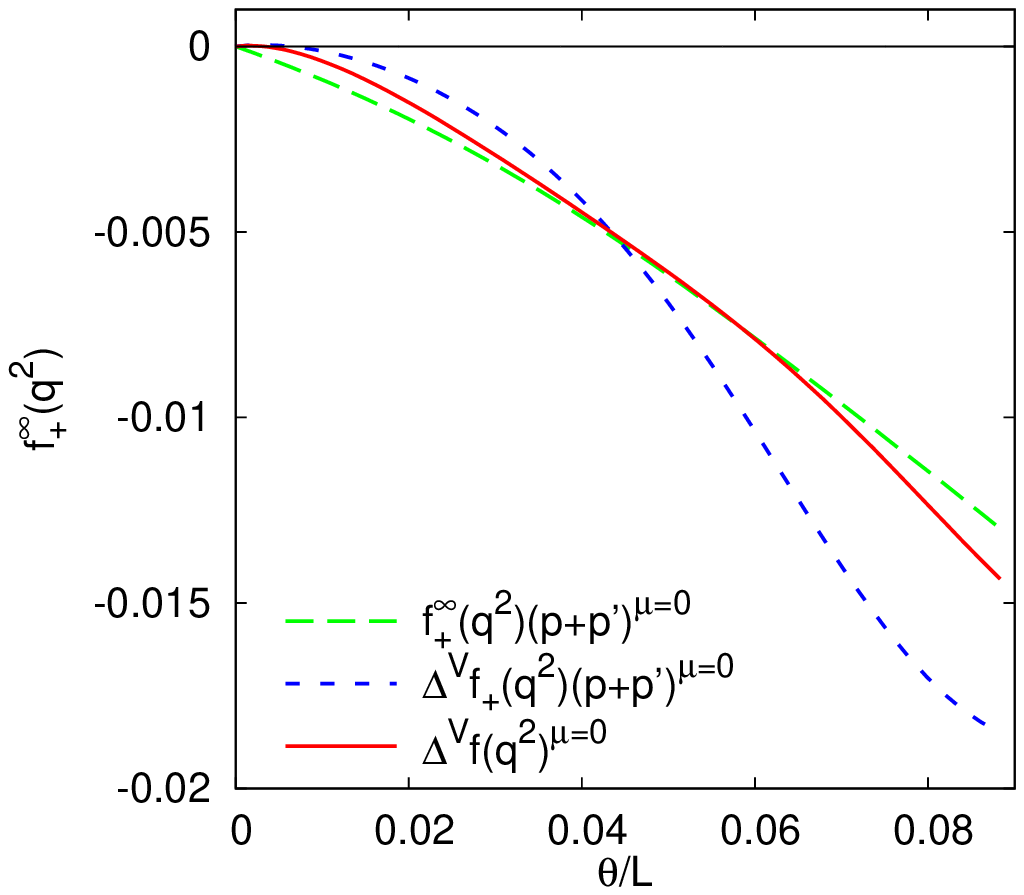}
    \hspace{0.05\textwidth}
    \includegraphics[width=\figWidth\textwidth]{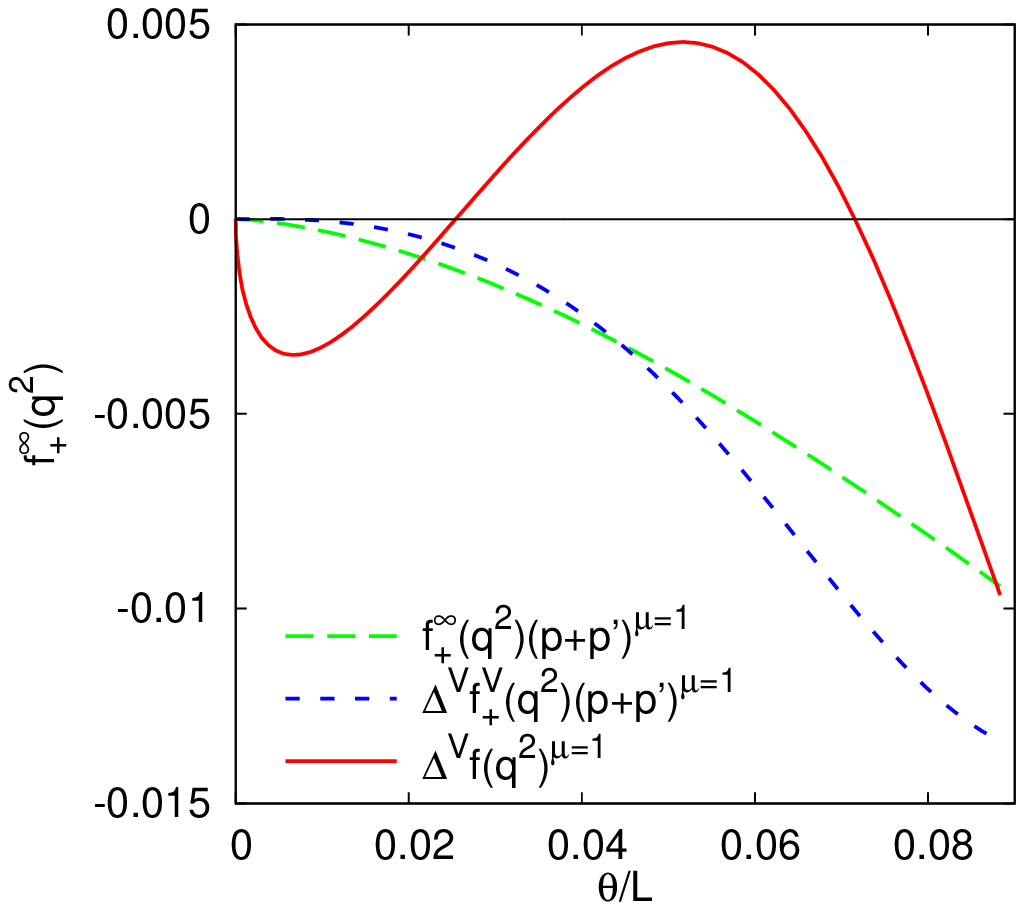}
  \end{center}
  \caption{\label{figfvcomp} Left: $\mu=0$ Right: $\mu=1$.
    One loop corrections to $f_+$. Plotted are the infinite volume correction, $f_+^\infty(q^2)$, the finite volume correction, $\Delta^V f_+$, and the full finite volume correction, $\Delta^V f^\mu = \Delta^V f_+(p+p^\prime)^\mu+ \Delta^V f_-q^\mu + \Delta^V h^\mu$. The one loop finite volume corrections are of similar size as the infinite volume one loop corrections.}
\end{figure}

We now split the form-factor as
\begin{align}
  f_\mu = \left(1+f^{\infty}_++\Delta^V f_+ \right)(p+p^\prime)_\mu
  +\Delta^V f_- q_\mu + \Delta^V h_\mu =
  -\frac{1}{\sqrt{2}} f_{\bar du\pi^+\pi^0\mu}\,.
\end{align}
where we have separated out the lowest order value of 1. In Fig.~\ref{figfvcomp} we plot the time and space components of the fifferent parts of $f_+$ as well as the full finite volume correction in a volume with size $m_\pi L = 2$. In order to look only at the loop contribution we have set $L_9^r = 0$. The renormalization scale is set to $\mu = 0.77$ GeV and we have used the following input constants
\begin{equation}
  \label{input1}
  m_\pi = 139.5~\mathrm{MeV}\,,\quad m_K = 495~\mathrm{MeV}\,,
  \quad m_\eta^2 = \frac{4}{3} m_K^2-\frac{1}{3}m_\pi^2,\quad
  F_\pi = 92.2~\mathrm{MeV}\,,\quad m_\pi L = 2\,.
\end{equation}
The momentum exchange $q^2$ is varied by changing the twist angle on the u-quark. From Fig.~\ref{figfvcomp} it is clear that the finite volume loop correction is of a similar size as the infinite volume loop correction.

\section{Decay constants}
Similar to the case of the electromagnetic form-factor, the decay constant also has a different momentum dependence with twisted boundary conditions than with periodic boundary conditions. We define the (axial-vector) decay constant in finite volume as 
\begin{align}
  \label{defFpi}
  \left< 0 |A_\mu^M | M(p) \right> = i\sqrt{2}F_M p_\mu + i\sqrt{2}F^V_{M\mu}\,,
\end{align}
where $M(p)$ is a meson and with the corresponding axial-vector decay constant
$A_\mu^M=\bar q \gamma_\mu\gamma_5({\lambda^M}/{\sqrt2}) q$. We also define a pseudo-scalar decay constant
\begin{equation}
  \left< 0|P^M|M(p)\right> = \frac{G_M}{\sqrt{2}}
\end{equation}
where $P^M=\bar q i\gamma_5(\lambda^M/\sqrt2)q$ is the pseudo-scalar current corresponding to the meson $M$. For flavour charged mesons composed of $\bar q q'$ the two matrix elements satisfy the Ward identity
\begin{equation}
  \partial^\mu \left< 0 |A_\mu^M | M(p) \right>
  = (m_q+m_{q'}) \left< 0|P^M|M(p)\right>
\end{equation}
which leads to 
\begin{equation}
  p^2 F_M+p^\mu F^V_{M\mu} = \frac{1}{2}(m_q+m_{q'})G_M\,.
  \label{eq:wardFG}
\end{equation}
When checking this Ward identity, it is important to use the correct momentum dependent mass.

\begin{figure}[t]
  \begin{center}
    \includegraphics[width=\figWidth\textwidth]{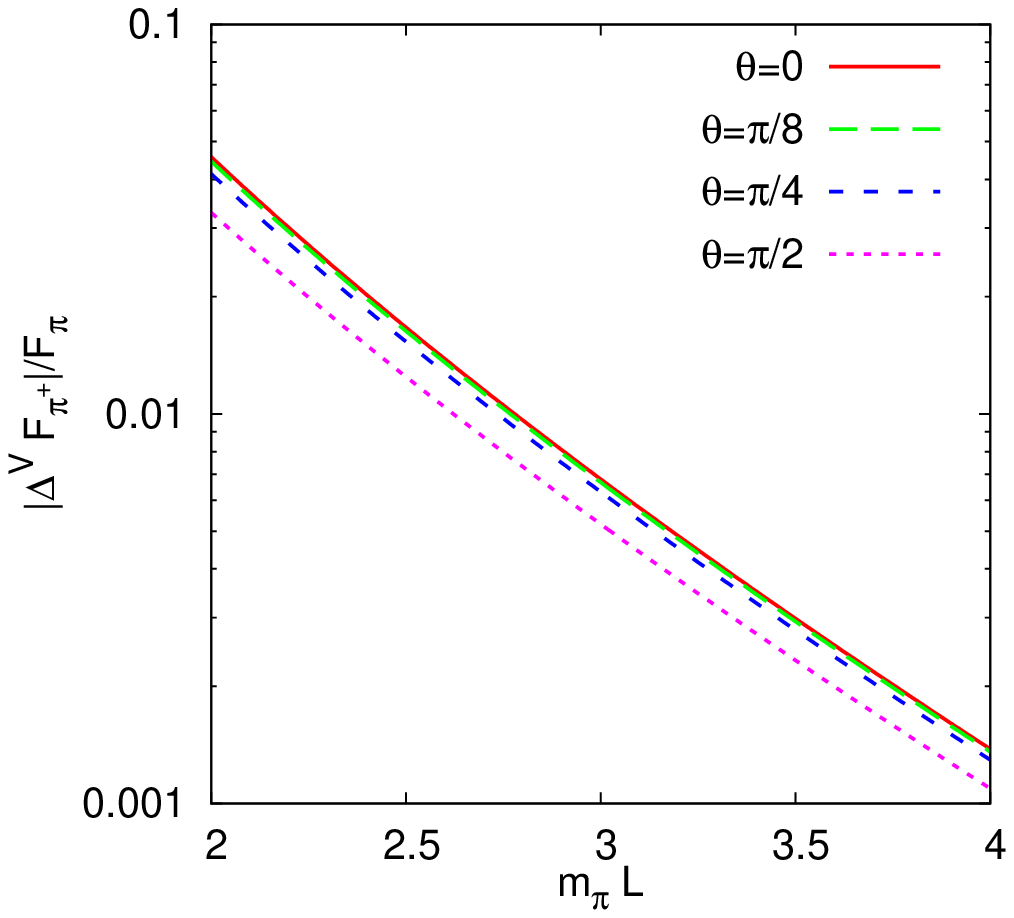}
    \hspace*{0.05\textwidth}
    \includegraphics[width=\figWidth\textwidth]{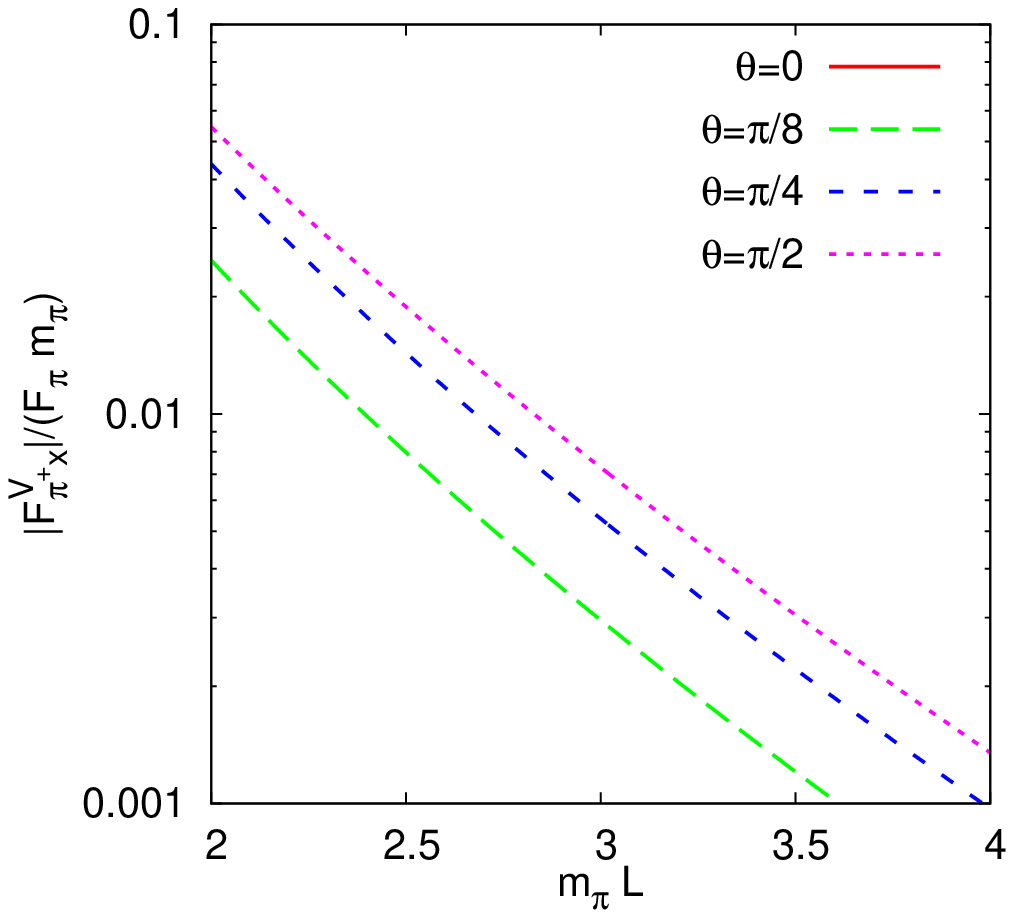}
  \end{center}
  \caption{\label{figDecay}
    The relative size of the finite volume correction to the decay constant of $\pi^+$ plotted against $m_\pi L$ for various twist angles on the u-quark. One the left we have plotted $\Delta^VF_M/F_\pi$ and on the right we have plotted $F^V_{Mx}/(F_\pi m_\pi)$.
  }
\end{figure}

The explicit expressions for $\pi^\pm$ decay are given by
\begin{align}
  \label{decayconstants}
  \Delta^V\! F_{\pi^\pm} &= \frac{1}{F_0}\left(
    \frac{1}{2} A^V(m_{\pi^+}^2) + \frac{1}{2} A^V(m_{\pi^0}^2) 
    + \frac{1}{4}A^V(m_{K^+}^2) + \frac{1}{4}A^V(m_{K^0}^2)\right)\,,
  \nonumber\\
  F^V_{\pi^\pm\mu} &=
  \pm\frac{1}{F_0}\left( 2 A^V_\mu(m_{\pi^+}^2) +  A^V_\mu(m_{K^+}^2) -  A^V_\mu(m_{K^0}^2)\right)\,,
  \nonumber\\
  \Delta^V\! G_{\pi^\pm}^V &= \frac{G_0}{F_0^2}
  \left(\frac{1}{2}A^V(m_{\pi^+}^2) + \frac{1}{4}A^V(m_{K^+}^2) 
    + \frac{1}{4}A^V(m_{K^0}^2) + \frac{1}{6}A^V(m_{\eta}^2)\right)\,.
\end{align}
The finite volume corrections to $\pi^+$ are plotted in Fig.~\ref{figDecay}.

As a final note on decay constants we note that due to the isospin breaking effect of twisted boundary conditions there is mixing in the decay of the flavour neutral mesons. We will not go into the details here but refer to \cite{Bijnens:2014yya} for the one loop results and \cite{ABT2} for how to treat mixing.

\section{Conclusions and outlook}
In this talk we have presented parts of our work on twisted boundary conditions using $SU(3)$ ChPT. The starting point is the observation that the odd tadpole integral is no longer zero when twisted boundary conditions are used. We note also that the broken lattice symmetry means that extra care is needed when defining quantities such as form-factors.

More specifically, we have given an introduction to twisted boundary conditions and how computations are affected. We have also presented results for mass, electromagnetic form-factor and decay constant for $\pi^\pm$. Finally we illustrated how the relevant Ward identities are satisfied. The procedure for checking Ward identities can be summarized as be careful and do things correctly! 

The size of the effects of twisted boundary conditions can be considered small in many cases, but depending on lattice size and the precision desired they can be non-negligible. 

We are now working on the one loop corrections to $Kl_3$ decay in partially twisted partially quenched staggered chiral perturbation theory, relevant for determining the $V_{us}$ mixing angle \cite{future}.

\acknowledgments
We thank the organizers for a very pleasant and well-run conference.
This work is supported in part by the Swedish Research Council grants
621-2011-5080 and 621-2013-4287.

\end{document}